\documentclass{PoS}

\usepackage{amsmath,amssymb,amscd,amsfonts,graphicx,bbm}

\rightline{\sffamily RBRC-778}\vspace*{-10mm}

\title{Perturbative $O(\alpha_s a)$ matching in static heavy
       and domain-wall light quark system}

\ShortTitle{Perturbative $O(\alpha_s a)$ matching in static heavy
            and DW light quark system}

\author{\speaker{Tomomi Ishikawa}\\
        RIKEN BNL Research Center, Brookhaven National Laboratory,
        Upton, New York 11973, USA\\
        E-mail: \email{tomomi@quark.phy.bnl.gov}}

\author{RBC and UKQCD Collaborations}

\abstract{
We discuss the perturbative $O(\alpha_s a)$ matching
in the static heavy and domain-wall light quark system.
The gluon action is the Iwasaki action and the link smearing is performed
in the static heavy action.
The chiral symmetry of the light quark realized by using the domain-wall
fermion formulation does not prohibit the mixing of the operators at $O(a)$.
The application of $O(a)$ improvement to the actual data shows
that the B meson decay constant $f_B$, the matrix elements ${\cal M}_B$ and
the B parameter $B_B$ have non-negligible effects,
while the effect on the SU(3) breaking ratio $\xi$ is small.
}

\FullConference{The XXVI International Symposium on Lattice Field Theory\\
		July 14-19 2008\\
		Williamsburg, Virginia, USA}

\begin{document}

\section{Introduction}

RBC/UKQCD Collaboration has been performing a large scale simulation
of the lattice QCD with dynamical domain-wall fermion (DWF)
\cite{Boyle:2007fn}.
In this project, we can intensively access the CKM matrix, which
includes b-quark physics.
To treat the b-quarks, the Heavy Quark Effective Theory (HQET)
\cite{Eichten:1989zv} is widely used.
The lattice calculation with the HQET, however, has following difficulties
(and solutions).
\begin{enumerate}
\item The static propagator is too noisy. ---
      This is basically because the static self-energy contains $1/a$ 
      power divergence.
      ALPHA collaboration investigated carefully this phenomena and
      introduced a modified static action which improves the signal to
      noise ratio \cite{Della Morte:2005yc}.
      The modification can be achieved by replacing the link variable
      in the static action with the smeared one, which is
      obtained by the 3-step hyper-cubic blocking.
      Using this the power divergence contributions in the static
      self-energy are largely reduced.

\item Non-perturbative matching with continuum is needed. ---
      If we include $O(1/m_b)$ correction in the HQET formulation,
      the continuum limit cannot be reached by using perturbative
      matching factor because of power divergence \cite{Heitger:2003nj}.
      Possible ways of non-perturbative matching are Schr\"{o}dinger
      functional scheme with step scaling technique and RI/MOM scheme.
\end{enumerate}
While the calculation can, in principle, be performed using the techniques
described above, the actual implementation is not easy.
There is an established way to apply the RI/MOM scheme for DW light quarks.
But it has not been applied to the HQET successfully.

As the first step of the project, the static approximation
(lowest order of the HQET) is valuable and an important approach to
the complete HQET.
In the static limit, the perturbative matching procedure is justified.
The perturbative matching at $O(\alpha_s)$ in the static heavy and
DW light quark system was calculated without link smearing
in \cite{Loktik:2006kz} and with link smearing in
\cite{Albertus:2007zz, Christ:2007cn}.
In this report, we present the calculation including the $O(a)$ improvement,
whose effect cannot be neglected in the heavy quark system that we are
considering here.

\section{Action setup}

We use the Iwasaki gluonic action and DW fermion with light quark mass
$m_q$ for the light quark sector.
The DWF has an optimized parameter $M_5$ which is called ``domain-wall
height'' and takes value $0<M_5<2$.
In the calculation of the matching factor, it is assumed that the
extension of the 5th dimension is infinity, which means the light quarks
have exact chiral symmetry.
For this sector, we do not carry out the link smearing.

For heavy quark sector, we use the static approximation with link
smearing:
\begin{eqnarray}
S_{\rm static}&=&
\sum_{\vec{x},t}\bar{h}(\vec{x},t)\left[
h(\vec{x},t)-W_0^{\dagger}(\vec{x},t-1)h(\vec{x},t-1)
\right],
\end{eqnarray}
where $h(\vec{x},t)$ is the effective heavy quark field and
$W_0(\vec{x},t)$ is the time-component of the smeared link variable.
If $W_0=U_0$, which is the original gauge link,
the action describes the one proposed by Eichten and Hill
~\cite{Eichten:1989kb}.
We use the 3-step hyper-cubic blocked link for $W_0$ with three
parameters $(\alpha_1, \alpha_2, \alpha_3)$.
Possible parameter choices are
\begin{eqnarray}
(\alpha_1, \alpha_2, \alpha_3)=\begin{cases}
(0.0 , 0.0, 0.0) & \hspace*{5mm} :{\rm unsmeared}~(W_0=U_0) \\
(1.0 , 0.0, 0.0) & \hspace*{5mm} :{\rm APE~with}~\alpha=1
~\cite{Albanese:1987ds}   \\
(0.75, 0.6, 0.3) & \hspace*{5mm} :{\rm HYP1}
~\cite{Hasenfratz:2001hp} \\
(1.0 , 1.0, 0.5) & \hspace*{5mm} :{\rm HYP2}
~\cite{Della Morte:2005yc}.
         		       \end{cases}
\end{eqnarray}

\section{$O(a)$ in the static heavy and light quark system}

In this report we mainly focus on the $O(a)$ improvement of operators,
and then we treat the matching factor between continuum HQET (CHQET)
and lattice HQET (LHQET).
Perturbative matching at one-loop between continuum QCD and CHQET was
obtained by Eichten and Hill~\cite{Eichten:1989zv}, which we can use.

\vspace*{-2mm}
\subsection*{\underline{Quark bilinear operator}}

We consider the on-shell $O(a)$ improved static heavy
$(h)$ - light $(q)$ quark bilinear 
\begin{eqnarray}
O_{\Gamma}^{\rm CHQET}&=&Z_{\Gamma}\left(1+b_{\Gamma}m_qa\right)\left[
O_{\Gamma}^{(0)}+c_{\Gamma}aO_{\Gamma}^{(1)}\right],
\label{EQ:Oa_bilinear}
\end{eqnarray}
relating the CHQET operator $O_{\Gamma}^{\rm CHQET}$ on the left hand side
and LHQET operators on the right hand side.
$O_{\Gamma}^{(0)}=\bar{h}\Gamma q$ and 
$O_{\Gamma}^{(1)}=\bar{h}\Gamma\vec{\gamma}\cdot\vec{D}q$
with
$\Gamma=\{1, \gamma_{\mu}, \gamma_5, \gamma_{\mu}\gamma_5, \sigma_{\mu\nu}\}$.
$Z_{\Gamma}$ is the overall matching factor between CHQET and LHQET,
$c_{\Gamma}$ and $b_{\Gamma}$ are the $O(pa)$ and $O(m_qa)$ improvement
coefficient, respectively.
In this expression, we reduced the dimension $4$ operators using the
equation of motions of static heavy and massless light quarks
\begin{eqnarray}
D_0h=0,\hspace*{10mm}
{\not\hspace*{-0.7mm}D}q=0.
\label{EQ:EOM}
\end{eqnarray}
The $O(pa)$ improvement of the heavy-light currents
with clover Wilson light quarks was investigated using one-loop perturbation
theory in non-relativistic QCD \cite{Morningstar:1997ep}
and the static approximation \cite{Ishikawa:1998rv}.
They showed that the $O(pa)$ effects give a large correction to 
the B meson decay constant $f_B$.
In the light-light quark system, the existence of chiral symmetry
guarantees the absence of $O(a)$ errors in the operators.
For the case of the static heavy-light quark system, however,
there are $O(a)$ effects even if we use chiral fermions for the light quarks.
This was already found in the clover Wilson light quark with Wilson
parameter $r=0$
(It is chirally symmetric, but there are doublers.) \cite{Ishikawa:1998rv}.

Now we consider the symmetries which the theory has.
In addition to the chiral symmetry in the light quark sector,
we have the heavy quark symmetry
$h\rightarrow e^{-i\phi_j\epsilon_{jkl}\sigma_{kl}}h$ for the heavy
quark sector.
These symmetries guarantee that $Z_{\Gamma}$ is independent on
$\Gamma$~\cite{Becirevic:2003hd}.
And also, $c_{\Gamma}=Gc$, $b_{\Gamma}=Gb$ with
$\gamma_0\Gamma\gamma_0=G\Gamma$, in which $c$ and $b$ are independent
on $\Gamma$.

\vspace*{-2mm}
\subsection*{\underline{Four-quark operator}}

We consider the four-quark operator ($\Delta B=2$)
which is relevant for the $B^0-\bar{B}^0$ mixing.
Its (full) QCD operator is
\begin{eqnarray}
O_L^{\rm QCD}=[\bar{b}\gamma_{\mu}^Lq][\bar{b}\gamma_{\mu}^Lq],
\end{eqnarray}
where $\gamma_{\mu}^L=\gamma_{\mu}P_L$ and also
$\gamma_{\mu}^R=\gamma_{\mu}P_R$.
At the one-loop level we need to take into account only the CHQET operator
\begin{eqnarray}
O_L=2[\bar{h}^{(+)}\gamma_{\mu}^Lq][\bar{h}^{(-)}\gamma_{\mu}^Lq],
\end{eqnarray}
for matching between the CHQET and LHQET operators,
where $h^{(+)}(h^{(-)})$ is the particle (anti-particle) of the static quark.
With the use of chiral fermions for the light quark,
the on-shell $O(a)$ improved four-fermion operator can be written in
\begin{eqnarray}
O_L^{\rm CHQET}
&=&Z_L\left[O_L+Z_L^{(1)}aO_{ND}+Z_L^{(m)}m_qaO_N\right],
\label{EQ:Oa_four-fermi}
\end{eqnarray}
where
\begin{eqnarray}
O_{ND}&=&
2[\bar{h}^{(+)}\gamma_{\mu}^Lq]
[\bar{h}^{(-)}\gamma_{\mu}^R(\gamma_i\vec{D}_i)q]
+4[\bar{h}^{(+)}P_Lq]
[\bar{h}^{(-)}P_R(\gamma_i\vec{D}_i)q]\nonumber\\
&&+2[\bar{h}^{(+)}\gamma_{\mu}^R(\gamma_i\vec{D}_i)q]
[\bar{h}^{(-)}\gamma_{\mu}^Lq]
+4[\bar{h}^{(+)}P_R(\gamma_i\vec{D}_i)q]
[\bar{h}^{(-)}P_Lq],\\
O_N&=&
2[\bar{h}^{(+)}\gamma_{\mu}^Lq][\bar{h}^{(-)}\gamma_{\mu}^Rq]
+4[\bar{h}^{(+)}P_Lq][\bar{h}^{(-)}P_Rq]\nonumber\\
&&+2[\bar{h}^{(+)}\gamma_{\mu}^Rq][\bar{h}^{(-)}\gamma_{\mu}^Lq]
+4[\bar{h}^{(+)}P_Rq][\bar{h}^{(-)}P_Lq],
\end{eqnarray}
$Z_L$ is an overall matching factor, $Z_L^{(1)}$ is the $O(pa)$
improvement coefficient and $Z_L^{(m)}$ is the $O(m_qa)$ improvement
coefficient.

\section{One-loop perturbative calculation of the $O(a)$ coefficients}

\vspace*{-2mm}
\subsection*{\underline{Quark bilinear operator}}

We calculate the matching factor and the $O(a)$ improvement coefficients
using one-loop perturbation theory.
The calculation is performed by comparing the light to heavy scattering
amplitude between the CHEQT and LHQET.
Now we consider the scattering amplitude with an initial light quark $q$
carrying momentum $p$ and a final heavy quark $h$ carrying momentum $k$.
In order to extract the on-shell $O(a)$ coefficients,
the amplitude is expanded in the external quark momenta $p$ and $k$ around
zero momentum and the light quark mass $m_q$ around zero mass.
Since the momenta obey the equation of motions (\ref{EQ:EOM}),
${\not\hspace*{-0.7mm}p}=0$ and $k_0=0$.
In the puerturbative calculation, we choose the Feynman gauge
and the UV divergences in the continuum calculation are regulated by
dimensional regularization and we use the $\overline{\rm MS}$ scheme
for the renormalization.
The IR divergences are regulated by introducing the gluon mass $\lambda$.

\begin{figure}[tbp]
\begin{center}
\includegraphics[scale=0.55, viewport = 0 0 700 230, clip]
                {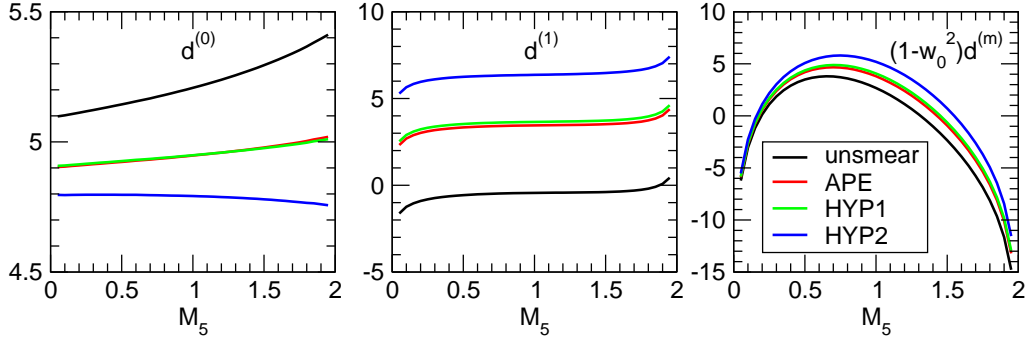}
\vspace*{-3mm}
\caption{Numerical values of $d^{(0)}$, $d^{(1)}$ and $(1-w_0^2)d^{(m)}$
         versus DW height $M_5$ for each link smearing.}
\label{FIG:numetical_values}
\end{center}
\vspace*{-5mm}
\end{figure}

The renormalized scattering amplitude for the CHQET at one-loop order
can be written in the form
\begin{eqnarray}
\langle h(k)|O_{\Gamma}|q(p)\rangle_{\rm cont}&=&
\left(1+\frac{\alpha_s}{4\pi}C_F{\cal A}_{\rm cont}^{(0)}\right)
\langle O_{\Gamma}^{(0)}\rangle_0
+\frac{\alpha_s}{4\pi}C_F{\cal A}_{\rm cont}^{(1)}
a\langle O_{\Gamma}^{(1)}\rangle_0\nonumber\\
&&\hspace*{39mm}+\frac{\alpha_s}{4\pi}C_F{\cal A}_{\rm cont}^{(m)}
m_qa\langle O_{\Gamma}^{(0)}\rangle_0,
\end{eqnarray}
where $\langle~~\rangle_0$ represents the tree level expectation value
of the amplitude, $C_F=(N_c^2-1)/(2N_c)$ with number of color $N_c$, and
\begin{eqnarray}
{\cal A}_{\rm cont}^{(0)}=
-\frac{3}{2}\ln\left(\frac{\lambda^2}{\mu^2}\right)+\frac{5}{4},\;\;\;
{\cal A}_{\rm cont}^{(1)}=
-\frac{8\pi}{3a\lambda},\;\;\;
{\cal A}_{\rm cont}^{(m)}=
-\frac{4\pi}{3a\lambda}.
\end{eqnarray}
In this expression $\mu$ is the renormalization scale parameter.
The scattering amplitude for the LHQET has the same IR divergence as
in the continuum:
\begin{eqnarray}
\langle h(k)|J_{\Gamma}|q(p)\rangle_{\rm latt}&=&
\left(1+\frac{\alpha_s}{4\pi}C_F{\cal A}_{\rm latt}^{(0)}\right)
\langle J_{\Gamma}^{(0)}\rangle_0
+\frac{\alpha_s}{4\pi}C_F{\cal A}_{\rm latt}^{(1)}
a\langle J_{\Gamma}^{(1)}\rangle_0\nonumber\\
&&\hspace*{40mm}+\frac{\alpha_s}{4\pi}C_F{\cal A}_{\rm latt}^{(m)}
(1-w_0^2)m_qa\langle J_{\Gamma}^{(0)}\rangle_0,
\end{eqnarray}
where $w_0=1-M_5$,
\begin{eqnarray}
{\cal A}_{\rm latt}^{(0)}=
-\frac{3}{2}\ln\left(a^2\lambda^2\right)+\frac{f+e_R}{2}+d^{(0)},\;\;\;
{\cal A}_{\rm latt}^{(1)}=
-\frac{8\pi}{3a\lambda}+d^{(1)},\;\;\;
{\cal A}_{\rm latt}^{(m)}=
-\frac{4\pi}{3a\lambda}+d^{(m)}.
\end{eqnarray}
The value of $f$ was obtained in \cite{Aoki:2002iq}.
Since we will use the fitting function $\sim e^{-Et}$, $e_R=e-\delta\hat{M}$,
the reduced value of $e$, is used \cite{Eichten:1989kb}.
The values are presented in Tab.~\ref{TAB:numerical_values}.
\begin{table}[tbp]
\begin{center}
\begin{tabular}{ccccc}
 \hline
                       & unsmeared  & APE     & HYP1    & HYP2     \\
 \hline
 $\delta\hat{M}$       & $12.979$ & $5.514$  & $4.910$  & $3.671$  \\
 $e$                   & $14.884$ & $1.429$  & $0.667$  & $-3.378$ \\
 $e_R=e-\delta\hat{M}$ & $1.906$  & $-4.085$ & $-4.243$ & $-7.049$ \\
 \hline
\end{tabular}
\caption{Numerical values of $\delta\hat{M}$, $e$ and $e_R$ for each link
         smearing.}
\label{TAB:numerical_values}
\end{center}
\vspace*{-5mm}
\end{table}
$d^{(0)}$, $d^{(1)}$ and $d^{(m)}$ are the finite parts of the vertex
correction whose values are shown in Fig.~\ref{FIG:numetical_values}.
After the matching we obtain the renormalized operator with $O(a)$ improvement
\begin{eqnarray}
O_{\Gamma}^{\rm CHQET}
&=&(1-w_0^2)^{-1/2}Z_w^{-1/2}
Z_{\Gamma}\left(1+b_{\Gamma}(1-w_0^2)m_qa\right)
\left[O_{\Gamma}^{(0)}+c_{\Gamma}aO_{\Gamma}^{(1)}\right],
\label{EQ:improved_current}
\end{eqnarray}
where
\begin{eqnarray}
Z_{\Gamma}&=&
1+\frac{\alpha_s}{4\pi}C_F\left[
\frac{3}{2}\ln\left(a^2\mu^2\right)
+\frac{5}{4}-\frac{f+e_R}{2}-d^{(0)}\right],
\label{EQ:Z_Gamma}\\
c_{\Gamma}&=&
-\frac{\alpha_s}{4\pi}C_FGd^{(1)},\;\;\;\;
b_{\Gamma}=
-\frac{\alpha_s}{4\pi}C_FGd^{(m)}.
\label{EQ:c_b_Gamma}
\end{eqnarray}
Because of our use of DW light quarks we need the DW-specific factors
$(1-w_0^2)=(1-(1-M_5)^2)$ and $Z_w=1+\frac{\alpha_s}{4\pi}C_Fz_w$
~\cite{Aoki:2002iq} in Eq. (\ref{EQ:improved_current}).
The $O(\alpha_s a)$ coefficients Eq. (\ref{EQ:c_b_Gamma})
are new results of this calculation.

Here we should briefly mention the $1/a$ power divergence in the operator
$O_{\Gamma}^{(1)}$ caused by the mixing with the lower dimensional operator 
$O_{\Gamma}^{(0)}$.
Since this power divergence is already in the $O(a)$ part,
the total contribution is $O(a^0)$ and we do not worry about it
in taking the continuum limit.
And also this $O(a^0)$ effect contributes at $O(\alpha_s^2)$,
which we can neglect in this one-loop calculation.
This is quite different from the power divergence that appears
in the $1/m_b$ expansion:
if the matching is done at $l$-th loop, the power divergence of
$\sim \alpha_s^{l+1}/a$ remains.
 
\vspace*{-2mm}
\subsection*{\underline{Four-quark operator}}

The calculation for the four-quark operator can be done
just by rearranging the quark bilinear results above.
After the matching we obtain the $O(a)$ improved operator
\begin{eqnarray}
O_L^{\rm CHQET}
&=&(1-(w_0)^2)^{-1}(Z_w)^{-1}Z_L
\left[O_L+Z_L^{(1)}aO_{ND}+Z_L^{(m)}(1-w_0^2)m_qaO_N\right].
\label{EQ:improved_four_quark}
\end{eqnarray}
As in the quark bilinear operator, we need the DW-specific factors
in Eq. (\ref{EQ:improved_four_quark}).
The matching factor and $O(a)$ coefficients are
\begin{eqnarray}
Z_L^{(0)}
&=&1+\frac{\alpha_s}{4\pi}\left[
4\ln\left(a^2\mu^2\right)
+\frac{7}{3}-\frac{10}{3}d^{(0)}-\frac{c}{3}-\frac{v}{3}
-\frac{4}{3}e_R-\frac{4}{3}f\right],\\
Z_L^{(1)}
&=&\frac{\alpha_s}{4\pi}\cdot 2d^{(1)},\;\;\;\;
Z_L^{(m)}
=\frac{\alpha_s}{4\pi}\cdot 2d^{(m)},
\end{eqnarray}
where the constant $v$ is the one-loop contribution from the diagram
in which the gluon connects light and light, which was obtained
in \cite{Aoki:2002iq}.
The constant $c$ arises when the gluon connects two heavy lines and is
given by $c=e_R$.

\section{Discussion}

In Eqs. (\ref{EQ:Oa_bilinear}) and (\ref{EQ:Oa_four-fermi}),
we used the operators $O_{\Gamma}^{(1)}$ and $O_{ND}$ which contain
covariant derivatives.
These operators, however, can be written in the form:
\begin{eqnarray}
\bar{h}^{(\pm)}\Gamma\vec{\gamma}\cdot\vec{D}q
&=&\mp G\partial_0\left(\bar{h}^{(\pm)}\Gamma q\right),\;\;\;\;
O_{ND}=2[\bar{h}^{(+)}\gamma_{\mu}^L q]
\left(\overleftarrow{\partial}_0-\overrightarrow{\partial}_0\right)
[\bar{h}^{(-)}\gamma_{\mu}^L q],
\label{EQ:total_derivative_op}
\end{eqnarray}
where we have used the equations of motion (\ref{EQ:EOM}).
This form is quite convenient
for taking the $O(a)$ improvement in correlation functions.
In the evaluation of the 2-point correlation function
$\langle A_0^{(-)\rm imp}(t)A_0^{(-)\dagger}(0)\rangle$,
where $A_0^{(-)}=\bar{h}^{(-)}\gamma_0\gamma_5q$,
we have
\begin{eqnarray}
\langle A_0^{(-)\rm imp}(t)A_0^{(-)\dagger}(0)\rangle&=&
\left(1+b_{\Gamma}(1-w_0^2)m_qa+c_AaE_{\rm bind}\right)
\langle A_0^{(-)}(t)A_0^{(-)\dagger}(0)\rangle.
\label{EQ:2-point}
\end{eqnarray}
$E_{\rm bind}$ is the binding energy of static heavy and light quark,
which is obtained in the correlator fitting.
Therefore, in order to accomplish the $O(a)$ improvement,
no further measurement is needed in the 2-point correlation function.
And also because the $O(m_qa)$ part can be neglected
due to its small size in many cases, we omit the $O(m_qa)$
in the following discussion.
Using the Eq.~(\ref{EQ:2-point}), we can evaluate the $O(a)$ improvement
of B meson decay constant $f_B$ like
\begin{eqnarray}
f_B^{\rm imp}
=\left(1+c_AaE_{\rm bind}\right)f_{B}.
\end{eqnarray}
For matrix element ${\cal M}_B$, B parameter $B_B$ and
SU(3) breaking ratio $\xi$ we use the vacuum saturate
approximation (VSA), and we obtain
\begin{eqnarray}
{\cal M}_B^{\rm imp}
\xrightarrow{\rm VSA}\left(1+2Z_L^{(1)}aE_{\rm bind}\right){\cal M}_B,\;\;\;
&&B_B^{\rm imp}
\xrightarrow{\rm VSA}\left(1+2(Z_L^{(1)}-c_A)aE_{\rm bind}\right)B_B,\nonumber\\
&&\xi^{\rm imp}
\xrightarrow{\rm VSA}
\left(1+Z_L^{(1)}a(E_{{\rm bind}(B_s)}-E_{{\rm bind}(B_d)})\right)\xi.
\end{eqnarray}

Now we roughly estimate these $O(\alpha_sa)$ effect using the actual
simulation data
($\beta=2.13$, $L^3\times T\times L_5=16^3\times 32\times 16$, $M_5=1.80$,
$m_{ud}a=\{0.01, 0.02, 0.03\}$, $m_sa=0.0359$)
which appeared in \cite{Albertus:2007zz}.
For this estimate the MF-improvement is taken into account.
In this case, $d^{(1)}=3.48({\rm APE}), ~6.41({\rm HYP2})$ and
$E_{\rm bind}\sim 0.6({\rm APE}), ~0.5({\rm HYP2})$.
The coupling constant has the range $\alpha_s\sim 0.15-0.35$, conservatively.
The conclusion is that the $O(\alpha_sa)$ effect of $f_B$ is
$3-8\%({\rm APE}), ~5-12\%({\rm HYP2})$, 
of ${\cal M}_B$ is $9-24\%({\rm APE}), ~15-36\%({\rm HYP2})$,
and of $B_B$ is $3-8\%({\rm APE}), ~5-12\%({\rm HYP2})$.
Using the assumption $(E_{{\rm bind}(B_s)}-E_{{\rm bind}(B_d)})\sim
(m_{B_s}-m_{B_d})$, the effect for $\xi$ is less than $2\%$.

\section{Summary}

We have presented a one-loop perturbative calculation of the $O(a)$
improvement coefficient for the static heavy - DW light quark system
taking into account the link smearing in the heavy quark sector.
Estimated $O(a)$ effect is not small in $f_B$, ${\cal M}_B$ and $B_B$,
but is small in $\xi$.
While perturbative matching has large ambiguities and its own limitations,
we deduce that this conclusions is not largely changed even in the
non-perturbative matching.\\

We thank all the member of the RBC and UKQCD Collaborations.

\end{document}